\documentclass[structabstract,longauth,letter]{aa}
\usepackage{graphicx}
\usepackage{txfonts}
\usepackage{natbib}
\bibpunct{(}{)}{;}{a}{}{,} 

\newcommand{\K}{\ensuremath{\,{\rm K}}}

\newcommand{\hii}{{\rm H\,}{{\sc ii}}}

\newcommand{\micron}{\mbox{$\mu$m}}
\begin{document}

\title{The physical properties of the dust in the RCW~120 H\,{\sc{ii}}
  region as seen by $Herschel$\thanks{{\it Herschel} is an ESA space
    observatory with science instruments provided by European-led
    Principal Investigator consortia and with important participation
    from NASA.}}

\author{L.D.~Anderson\inst{1} \and A.~Zavagno\inst{1} \and J.A.~Rod\'on\inst{1} \and D.~Russeil\inst{1} \and
 A.~Abergel\inst{2} \and
 P.~Ade\inst{3} \and
 P.~Andr\'e\inst{4} \and
 H.~Arab\inst{2} \and
 J.-P.~Baluteau\inst{1} \and
 J.-P.~Bernard\inst{6} \and
 K.~Blagrave\inst{13} \and
 F.~Boulanger\inst{2} \and
 M.~Cohen\inst{8} \and
 M.~Compiegne\inst{13} \and
 P.~Cox\inst{9} \and
 E.~Dartois\inst{2} \and
 G.~Davis\inst{10} \and
 R.~Emery\inst{16} \and
 T.~Fulton\inst{19} \and
 C.~Gry\inst{1} \and
 E.~Habart\inst{2} \and
 M.~Huang\inst{10} \and
 C.~Joblin\inst{6} \and
 S.C.~Jones\inst{15} \and
 J.~Kirk\inst{3} \and
 G.~Lagache\inst{2} \and
 T.~Lim\inst{16} \and
 S.~Madden\inst{4} \and
 G.~Makiwa\inst{15} \and
 P.~Martin\inst{13} \and
 M.-A.~Miville-Desch\^enes\inst{2} \and
 S.~Molinari\inst{14} \and
 H.~Moseley\inst{18} \and
 F.~Motte\inst{4} \and
 D.A.~Naylor\inst{15} \and
 K.~Okumura\inst{4} \and
 D.~Pinheiro Gocalvez\inst{13} \and
 E.~Polehampton$^{15,16}$ \and
 P.~Saraceno\inst{14} \and
 S.~Sidher\inst{16} \and
 L.~Spencer\inst{15} \and
 B.~Swinyard\inst{16} \and
 D.~Ward-Thompson\inst{3} \and
 G.J.~White\inst{16,21}
  }

   \institute{
 Laboratoire d'Astrophysique de Marseille (UMR\,6110 CNRS \& Universit\'e de Provence), 38 rue F.  
 Joliot-Curie, 13388 Marseille Cedex 13, France \and
 Institut dÕAstrophysique Spatiale, UMR\,8617, CNRS/Universit\'e Paris-Sud\,11, 91405 Orsay, France  \and
 Department of Physics and Astronomy, Cardiff University, Cardiff, UK \and
 CEA, Laboratoire AIM, Irfu/SAp, Orme des Merisiers, F-91191 Gif-sur-Yvette, France \and
 Dept. of Physics \& Astronomy, University College London, Gower Street, London WC1E 6BT, UK \and
Centre d'\'etudes spatiales des rayonnements (CESR), Universit\'e de Toulouse (UPS), CNRS, UMR\,5187, 9 avenue du colonel Roche, 31028 Toulouse cedex 4, France \and
 CNRS/INSU, Laboratoire dÕAstrophysique de Bordeaux, UMR\,5804, BP 89, 33271 Floirac cedex, France \and
 Radio Astronomy Laboratory, University of California, Berkeley, USA \and
 IRAM, Grenoble, France \and
 National Astronomical Observatories (China) \and
National Research Council of Canada, Herzberg Institute of Astrophysics, Victoria, Canada \and
 Department of Physics and Astronomy, University of British Columbia, Vancouver, Canada \and
 Canadian Institute for Theoretical Astrophysics, Toronto, Ontario, M5S 3H8, Canada  \and
 Istituto di Fisica dello Spazio Interplanetario, INAF, Via del Fosso del Cavaliere 100, I-00133 Roma, Italy \and
 Institute for Space Imaging Science, University of Lethbridge, Lethbridge, Canada \and
 Space Science Department, Rutherford Appleton Laboratory, Chilton, UK \and
 Centre for Astrophysics and Planetary Science, School of Physical Sciences, University of Kent, Kent, UK \and
  NASA - Goddard SFC, USA \and
 Blue Sky Spectrosocpy Inc, Lethbridge, Canada \and
Laboratoire des Signaux et Systmes, SUPELEC , Plateau de Moulon, 91192 Gif-sur-Yvette Cedex, France \and
Department of Physics \& Astronomy, The Open University, Milton Keynes MK7 6AA, UK}

\date{Received / Accepted}

\abstract{RCW~120 is a well-studied, nearby Galactic \hii\ region with
  ongoing star formation in its surroundings.  Previous work has shown
  that it displays a bubble morphology at mid-infrared wavelengths,
  and has a massive layer of collected neutral material seen at sub-mm
  wavelengths.  Given the well-defined photo-dissociation region (PDR)
  boundary and collected layer, it is an excellent laboratory to study
  the ``collect and collapse'' process of triggered star formation.
  Using ${\it Herschel}$~Space~Observatory data at 100, 160, 250, 350,
  and 500\,\micron, in combination with ${\it Spitzer}$ and
  APEX-LABOCA data, we can for the first time map the entire spectral
  energy distribution of an \hii\ region at high angular resolution.}{
  We seek a better understanding of RCW~120 and its local environment
  by analysing its dust temperature distribution.  Additionally, we
  wish to understand how the dust emissivity index, $\beta$, is
  related to the dust temperature.}{We determine dust temperatures in
  selected regions of the RCW~120 field by fitting their spectral
  energy distribution (SED), derived using aperture photometry.
  Additionally, we fit the SED extracted from a grid of positions to
  create a temperature map.}{We find a gradient in dust temperature,
  ranging from $\gtrsim\,30$\,K in the interior of RCW~120, to
  $\sim\,20$\,K for the material collected in the PDR, to
  $\sim\,10$\,K toward local infrared dark clouds and cold filaments.
  There is an additional, hotter ($\sim100$\,K) component to the dust
  emission that we do not investigate here.  Our results suggest that
  RCW~120 is in the process of destroying the PDR delineating its
  bubble morphology.  The leaked radiation from its interior may
  influence the creation of the next generation of stars.  We find
  support for an anti-correlation between the fitted temperature and
  $\beta$, in rough agreement with what has been found previously.
  The extended wavelength coverage of the ${\it Herschel}$ data
  greatly increases the reliability of this result.}{}
 \maketitle

\titlerunning{Dust Properties of RCW~120}
\authorrunning{Anderson et al.}

\section{Introduction}
RCW~120 \citep{rcw60} is a Galactic \hii\ region that displays a ring
morphology at mid-infrared and sub-mm wavelengths, and is presumably a
bubble viewed in projection.  It has recently been studied by
\citet[][hereafter ZA07]{zavagno07} and \citet[][hereafter
  DE09]{deharveng09} in the context of triggered star formation.  It
is only 1.3\,kpc from the Sun (see ZA07, and references therein), and
is thus one of the closest Galactic \hii\ regions.

ZA07 analysed the 1.2\,mm emission of RCW~120 and found a fragmented
layer of neutral material adjacent to the photo-dissociation region
(PDR).  They identified eight millimeter condensations, five of which
lie in the collected layer of material, but found no massive young
stellar objects (YSOs) within the condensations.  They did, however,
locate numerous YSOs surrounding RCW~120, indicating that star
formation is active in the region.  Using APEX-LABOCA observations at
870\,\micron, and {\it Spitzer}-MIPS observations at 24\,\micron, DE09
extended this work and calculated column densities and masses for the
sub-mm condensations (plus an additional condensation, \#9).  They
found additional YSOs in the field, including a chain of 11
evenly-spaced YSOs inside the most massive condensation and a very
dense core harboring a (possibly class 0) YSO.  This work highlights
the impact RCW~120 is having on star formation far from the ionizing
source.  An analysis of the YSOs in the RCW~120 field using the
{\it Herschel} data presented here is given in a companion paper \citep{zavagno10}.

The ring morphology shown by RCW~120 is a common feature of Galactic
\hii\ regions; \citet{churchwell06, churchwell07} identified almost
600 such objects in the {\it Spitzer}-GLIMPSE data \citep{benjamin03}.
\citet{deharveng10} have shown that over 85\% of infrared (IR) bubbles
enclose \hii\ regions.  Because of their morphology, it is easy to
locate the PDRs of such bubbles and one
can more easily identify the swept-up material that is necessary for
the collect and collapse process \citep{elmegreen77}.  These objects
present an opportunity to assess the efficiency of triggered star
formation throughout the entire Galaxy.  Data from the {\it Herschel} telescope
\citep{pilbratt10} allow us for the first time to map the dust
temperature variations over an entire \hii\ region at high resolution.
We can thus better determine the effect \hii\ regions have on the
creation of the next generation of stars.

\section{Data}
RCW~120 was observed by the {\it Herschel} telescope on 9 October 2009
with the PACS \citep{poglitsch10} and SPIRE \citep{griffin10}
as part of, respectively, the HOBYS \citep{motte10} and
``Evolution of Interstellar Dust" \citep{abergel10} guaranteed
time key programs.  Data were taken in five wavelength bands: 100 and
160\,\micron\ for PACS (at resolutions of $10\arcsec$ and $14\arcsec$
and a final map size of $30\arcmin\times30\arcmin$), and 250, 350, and
500\,\micron\ for SPIRE (at resolutions of $18\arcsec$, $25\arcsec$,
and $36\arcsec$ and a final map size of $22\arcmin\times22\arcmin$).  We
reduced these data using HIPE version 2.0.  The SPIRE images used here
are level 2 products, produced by the SPIRE photometer pipeline.  We
reduced the PACS data using a script provided by M. Sauvage.  Recent
calibration changes for the PACS and SPIRE data were taken into
account \citep{swinyard10}; throughout we use calibration
uncertainties of 10\% and 20\% respectively for the PACS
100\,\micron\ and 160\,\micron\ bands, and 15\% for all three SPIRE
bands \citep{swinyard10}.

\begin{figure}

\resizebox{\hsize}{!}{\includegraphics[scale = 0.8]{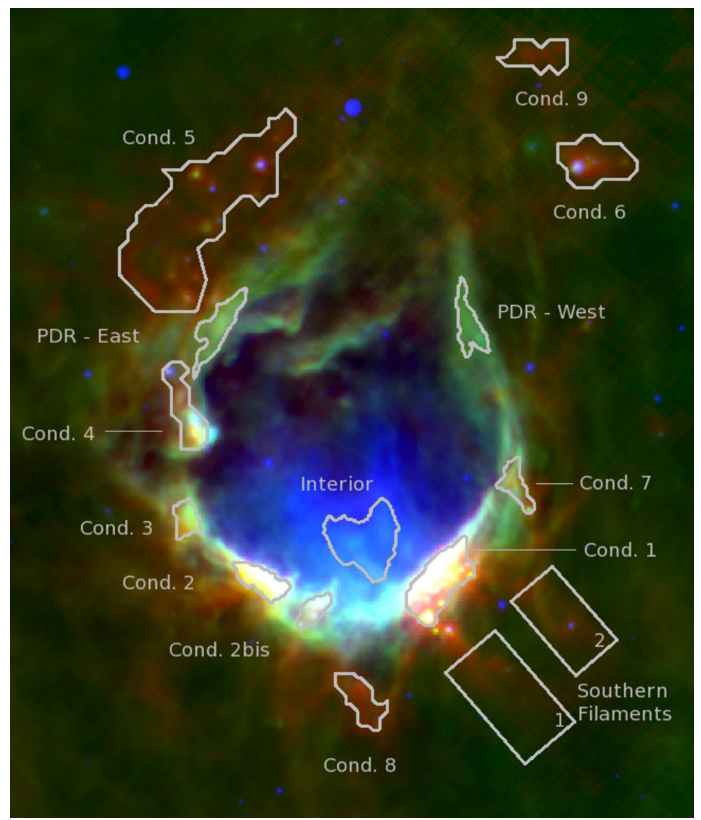}}
    \caption{Three-color image of RCW~120 composed of
      250\,\micron\ (red) and 100\,\micron\ (green) {\it Herschel}
      data, and 24\,\micron\ MISPGAL data (blue).  The entire field is
      $18\arcmin \times 22\arcmin$.  Regions used for aperture
      photometry are shown and labeled.  The coldest dust appears red,
      and is contained mostly in filaments on the southern edge of the
      bubble and within the infrared dark clouds to the north.  The
      hottest dust appears blue and is largely confined to the
      interior region of the bubble, coincident with the ionized
        gas.  The interior of the bubble has little emission at
      wavelengths $\ge250$\,\micron.}
    \label{fig:apphot}
\end{figure}

We also utilize {\it Spitzer}-MIPSGAL data \citep{carey09} at
24\,\micron\ and 70\,\micron\ (at resolutions of $6\arcsec$ and
$18\arcsec$), and APEX-LABOCA data at 870\,\micron\ from D09 (at a
resolution of $19\arcsec$).  The MIPSGAL 70\,\micron\ maps have strong
striping that we mediate with median filtering of the scan-legs
\citep[see][]{gordon07}.  In Fig.~\ref{fig:apphot} we show a
$18\arcmin \times 22\arcmin$ three-color image with SPIRE
250\,\micron\ (red), PACS 100\,\micron\ (green), and
MIPSGAL 24\,\micron\ data (blue).  Regions of interest are
superimposed on the data; the condensation numbers are from ZA07 and
DE09.

\section{Dust properties}

We assume the dust emission in RCW~120 can be modeled by an optically
thin grey-body, and that the emissivity of the dust grains can be fitted
with a power law \citep{hildebrand83}:
\begin{equation}
S_\nu \propto \Omega B_\nu(T) \, \kappa_0 \left(\frac{\nu}{\nu_0}\right)^{\beta} N,
\label{eq:grey}
\end{equation}
where $S_\nu$ is the measured flux density at frequency $\nu$, $\Omega$ is the
observing beam solid angle, $B_\nu(T)$ is the Planck function for
temperature $T$ at frequency $\nu$, $\kappa_0 (\nu/\nu_0)^\beta$ is
the dust opacity, $\beta$ is the dust emissivity index, and $N$ is the
dust column density.  Each line of sight has variations in the dust
temperature and emissivity; the derived temperatures and emissivities
therefore represent average values, weighted by the strongest emitting
components.


There has been much discussion about the value of the dust emissivity
index, $\beta$, and whether its value is anti-correlated with the dust
temperature, $T$.  Generally, $\beta$ is thought to range from 1 to 2, but
may also vary with wavelength \citep{meny07}.
\citet{dupac03} found that $\beta$ and $T$ are inversely related.
\citet{desert08} found a similar inverse-relation
to exclude a constant value of $\beta$ at the 99.9\% confidence level.
\citet{shetty09a, shetty09b}, however, suggested that this
relationship arises due to the influence of noise in a least-squares
fit of Eq.~\ref{eq:grey}, and from the combination of multiple
emission components along the line of sight.
A correlation between $\beta$ and $T$ may indicate a change in
  dust properties at high density \citep[][]{stepnik03}.

\subsection{Aperture photometry \label{sec:apphot}}
To determine the dust temperature structure of RCW~120, we performed
aperture photometry measurements on selected areas in the RCW~120
field.  We resampled all images to the resolution of the MIPSGAL
24\,\micron\ data to avoid pixel edge effects.  There are three
different areas of interest in the field of RCW~120: the interior of
the bubble, the PDR (including
condensations), and local infrared dark clouds (IRDCs) that appear in
emission at the PACS and SPIRE wavelengths.  The apertures we used are
shown in Fig.~\ref{fig:apphot}.  For each aperture, we selected a
nearby background aperture that best characterizes the local
background emission.

For all apertures, we fit a single temperature to the coldest emission
component of the spectral energy distribution (SED) using a
  non-linear least squares routine in two trials: once with
$\beta=2$, and once with $\beta$ allowed to vary.  For all apertures,
the 24\,\micron\ emission is significantly greater than what would be
predicted by a single, cold temperature.  This hotter component,
likely caused in part by transiently heated small grains, is not
well-constrained by the available data and therefore we ignore it in
the present work.  We also exclude the 70\,\micron\ data point
  (and sometimes the 100\,\micron\ data point) if it is inconsistent
  with the emission from a single, cold temperature component.  We leave the column density as a free parameter.

\begin{figure}[!ht]

  \resizebox{\hsize}{!}{\includegraphics[scale =0.8]{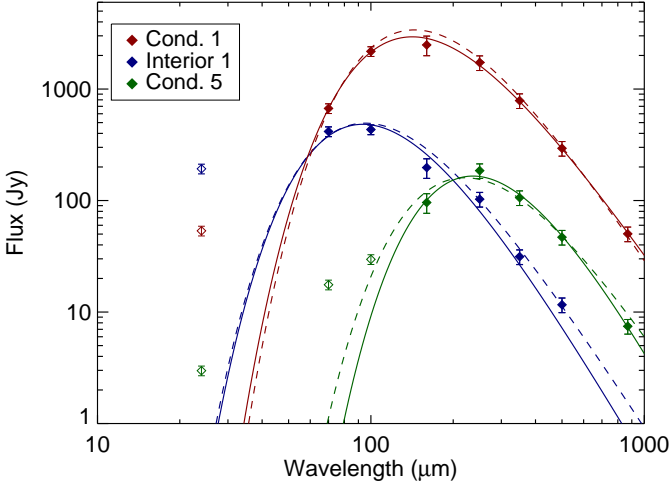}}
  
  \caption{Example SED fits for the ``Condensation 1'' (red),
    ``Interior 1'' (blue) and ``Condensation 5'' (green) apertures.
    Points included in the SED fits are shown filled; points excluded
    from the fits are shown open.  The fit with $\beta$ allowed to
    vary is shown with a solid line and the fit with $\beta=2$ is
    shown with a dashed line.  Apparent is the difference in
    temperature between the three apertures, and the higher
    temperature dust detected at 24\,\micron\ (and 70\,\micron\ and
    100\,\micron\ for Condensation 5).}

  \label{fig:sed_examples}
\end{figure}


\begin{table}[!ht]
\caption{Dust properties derived from aperture photometry}              
\label{tab:apphot}      
\centering                           
\begin{tabular}{lccc}        
\hline\hline                 
\multicolumn{1}{|l|}{Name} & \multicolumn{1}{|c|}{$T$\,(K)} & $T$\,(K) & \multicolumn{1}{|c|}{$\beta$} \\
\multicolumn{1}{|c|}{ } & $\beta = 2$ & \multicolumn{2}{|c|}{$\beta$ free} \\
\hline                        
Interior &  $  30.4\pm   0.9$  &  $  28.9\pm   2.0$  &  $   2.4\pm   0.2$  \\
PDR - West &  $  27.1\pm   0.6$  &  $  20.5\pm 1.0$  &  $   2.9\pm   0.2$  \\
PDR - East &  $  23.1\pm   0.7$  &  $  22.9\pm 2.1$  &  $   2.1\pm   0.2$  \\
Cond. 1 &  $  20.1\pm   0.3$  &  $  21.6\pm 1.1$  &  $   1.7\pm   0.2$  \\
Cond. 2 &  $  22.2\pm   0.4$  &  $  22.1\pm 1.2$  &  $   2.1\pm   0.2$  \\
Cond. 3 &  $  21.5\pm   0.3$  &  $  22.2\pm 1.2$  &  $   2.0\pm   0.2$  \\
Cond. 4 &  $  21.0\pm   0.4$  &  $  23.0\pm 1.2$  &  $   1.8\pm   0.2$  \\
Cond. 5 (IRDC) &  $  12.6\pm   0.5$  &  $  \phantom{0} 9.2\pm   0.9$  &  $   3.2\pm   0.4$  \\
Cond. 6 (IRDC) &  $  13.1\pm   0.5$  &  $  10.5\pm   1.2$  &  $   2.8\pm   0.4$  \\
Cond. 7 &  $  21.8\pm   0.4$  &  $  22.7\pm 1.2$  &  $   1.9\pm   0.2$  \\
Cond. 8 &  $  14.3\pm   0.6$  &  $  \phantom{0} 9.9\pm 1.0$  &  $   3.3\pm   0.4$  \\
Cond. 9 (IRDC) &  $  12.8\pm   0.5$  &  $  10.2\pm   1.1$  &  $   2.9\pm   0.4$  \\
Cond. 10 &  $  23.5\pm   0.7$  &  $  25.9\pm  2.8$  &  $   1.9\pm   0.2$  \\
Southern Filaments 1 &  $  14.6\pm   0.7$  &  $  10.7\pm   1.3$  &  $   3.0\pm   0.4$  \\
Southern Filaments 2 &  $  13.5\pm   0.6$  &  $   \phantom{0} 9.6\pm   1.0$  &  $   3.2\pm   0.4$  \\
\hline                                    
\end{tabular}
\end{table}

The results of our aperture photometry are shown in
Table~\ref{tab:apphot}, and examples of the SED fits are shown in
Fig.~\ref{fig:sed_examples}.  Uncertainties listed in
Table~\ref{tab:apphot} are the formal $1\sigma$ values, taking into
account uncertainties in the photometry measurements and calibration
uncertainties.  The direction of the interior of RCW~120
has dust temperatures of $\gtrsim 30\K$.  This result is rather
uncertain, however, because of the lack of emission above the background level in the interior of RCW~120 at wavelengths
$\ge 250$\,\micron.
The lack of emission at long wavelengths is evidence for a flattened,
two dimensional ring instead of a bubble, as suggested by
\citet{beaumont10}.
The condensations and PDR have characteristic temperatures of
$\sim20$\,K and the IRDCs have temperatures of $\sim 10$\,K, on the
low end of the temperature distribution for IRDCs found by \citet{peretto10}.  The temperatures change little in the two trials.
Uncertainties in temperature are small (often $\lesssim 5\%$), and errors in $\beta$ are
generally 10\%.

\begin{figure}

  \resizebox{\hsize}{!}{\includegraphics[scale = 0.8]{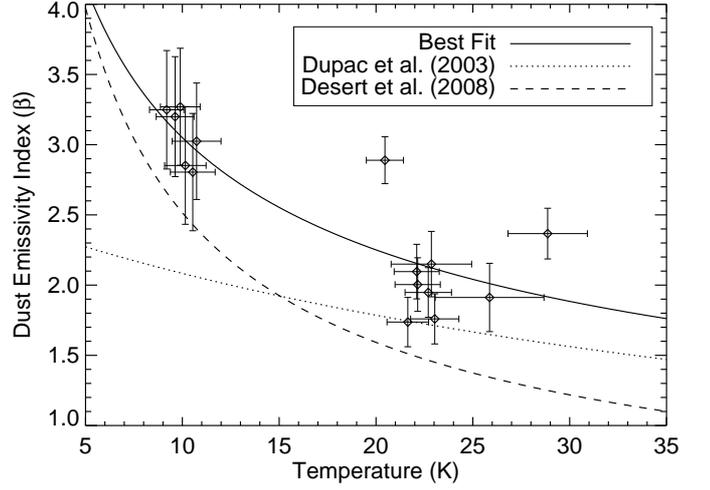}}
  
  \caption{The relationship between the fitted temperature and dust
    emissivity index $\beta$.  The points are the results of our
    aperture photometry measurements.  The solid curve shows the
    best-fit line, the dotted curve shows the relationship found by
    \citet{dupac03}, and the dashed curve shows the relationship found
    by \citet{desert08}.}
  
  \label{fig:t_vs_beta}
\end{figure}

The $\beta$ and $T$ values, together with the empirically derived
curves of \citet{dupac03} and \citet{desert08}, are shown in
Fig.~\ref{fig:t_vs_beta}.  Figure~\ref{fig:t_vs_beta} clearly shows
two $T$-$\beta$ groups, one for the PDR (and the condensations along
the PDR) at $(T, \beta) \simeq (20, 2.0)$, and one for dark clouds and
cold filaments at $(T, \beta) \simeq (10, 3.0)$.  We find an
anti-correlation between $\beta$ and the dust temperature.  Fitting a
regression line of the same functional form as that of
\citet{desert08} reveals:
\begin{equation}
\beta = (8.4\pm1.9) \times T^{-0.4\pm0.1}\,.
\end{equation}
The functional form of the relationship given in \citet{dupac03} does
not fit our data.  While these data do support the $\beta-T$
relationship, this result should be interpreted with caution because
the calibration for the {\it Herschel} instruments is not yet
finalized.  Also, the fits to many of the cold filaments are consistent
with a $\beta$ value of two.  For example, the $\beta=2$ fit for
Condensation \#5 is also acceptable as it falls roughly within the
formal errors (see Fig.~\ref{fig:sed_examples}).

\subsection{Temperature map}
While aperture photometry is well-suited for investigating the average
dust properties of small areas in the field of RCW~120, a higher
resolution map of dust temperature is useful for determining the
small-scale variations in dust properties.  To construct such a
temperature map, we fit the SED extracted at each pixel with a
single-temperature grey-body model (Eq.~\ref{eq:grey}), again
leaving the column density as a free parameter.
\citet{schnee07} showed that when attempting to simultaneously
constrain $\beta$ and $T$, uncertainties in the derived parameters can
be very large due to the effect of noise in the data.  We find that
this is indeed the case and
therefore in the following we fix $\beta$ to a value of two,
  which is valid for most areas of the map (see Table
  \ref{tab:apphot}).

Fitting the SED of each pixel requires matching the resolution and
pixel locations of all wavelengths to those of the lowest resolution
image.  The SPIRE 500\,\micron\ data point has little impact on the
derived fit parameters because we have higher-resolution data at
870\,\micron; we exclude the 500\,\micron\ data point so we can rebin
to the higher resolution of the SPIRE 350\,\micron\ data instead.  We
smooth all images with a two-dimensional Gaussian representing the
SPIRE beam at 350\,\micron, and rebin to the 350\,\micron\ resolution
using Montage\footnote{http://montage.ipac.caltech.edu/}.  We subtract
an average background value from the data at all wavelengths,
estimated from a field nearly devoid of emission.  For all but the
coldest dust, the {\it Spitzer} 70\,\micron\ data point is dominated
by emission from the cold component; we use this data point in the SED
fits here.  The inclusion of these data, however, causes the fitted
temperature to be overestimated in very cold regions (see fits in
Fig.~\ref{fig:sed_examples}).

\begin{figure}

\resizebox{\hsize}{!}{\includegraphics[scale = 0.4]{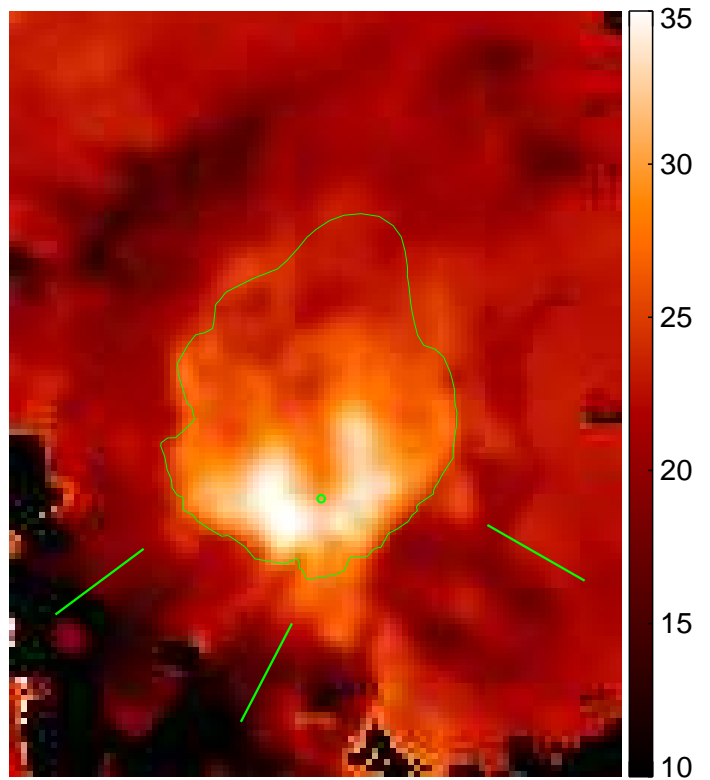}}

\caption{The temperature map derived from the SED fits of each pixel,
  over the same area as Fig.~\ref{fig:apphot}.  When deriving these
  maps, we fixed $\beta$ to a value of 2.  The interior edge of the
  PDR, found by tracing the {\it Spitzer} 8.0\,\micron\ data, is shown
  as the green line.  The location of the ionizing star is shown with
  the small green circle.  The location of hotter dust outside the
  ionization front is shown with the three green lines.}
  
\label{fig:sedmap}
\end{figure}

The temperature map is shown in Fig.~\ref{fig:sedmap}, for the same
area shown in Fig.~\ref{fig:apphot}.  We find the same results found
with the aperture photometry method: the interior of RCW~120 is $\sim
30$\,K the PDR is $\sim20$\,K, and the dark clouds are $\sim10$\,K.
The highest temperatures (up to $\sim 40$\,K) are found toward the
southern edge of the bubble, surrounding the location of the main
ionizing star.  Errors in the derived temperatures for individual
pixels are generally $<20\%$.  The exact location of the ionizing
source is devoid of hot dust, an effect that may be caused by the
action of stellar winds.  We note, however, that the ionizing star of
RCW120 has been found to have relatively weak winds \citep{martins10}.

The action of the ionized material appears to be creating openings in
the PDR in several locations, shown with green lines in
Fig.~\ref{fig:sedmap}.  These openings can also be seen in
Fig.~\ref{fig:apphot}.  There appears to be dust that is hotter by
$\sim\,5$\,K outside the PDR at these locations, heated by radiation
leaking through the holes in the PDR.  This higher temperature
dust may lead to the collapse of future generations of stars as it
interacts with local molecular material outside the PDR.  For the
opening due south, this effect may have aided the creation of
the sources inside Condensation~\#8 seen in Fig.~\ref{fig:apphot};
the directions of the other openings show no indication of
increased star-formation activity.

There are a number of cold filaments in the field of RCW~120 that have
dust temperatures of $\sim 10$\,K.  All such cold filaments appear
bright at wavelengths $>250$\,\micron\ and therefore appear red in
Fig.~\ref{fig:apphot}.  These filaments are associated with IRDCs
seen in {\it Spitzer} 8.0\,\micron\ data.  The most obvious cold
patches in the field are from the IRDCs to the north, but there are
numerous, thin filaments throughout the field.  A number of cold
filaments are oriented radially away from RCW~120, the coldest of
which are towards the south-west.  We suggest that these southern
filaments are shaped by radiation from RCW~120 leaving the ionized
region.  We suggest that the boundary of RCW~120 does not entrap all
the emission; where the radiation leaks out, the filaments
are compressed into the radial segments we see.  In the direction of
large condensations, the radiation is confined to the bubble area.

\section{Conclusions}
We have analysed the dust properties of the nearby bubble \hii\ region
RCW~120 using aperture photometry and temperature maps of {\it
  Herschel}, {\it Spitzer}, and APEX-LABOCA data.  We have found a
gradient of dust temperature, from $\sim 10$\,K for local infrared
dark clouds, to $\sim20$\,K for the PDR (including sub-mm)
condensations, to $\gtrsim 30\,K$ for the interior of the bubble.

Our results show support for a power-law form of the anti-correlation
between the dust temperature $T$ and the dust emissivity index
$\beta$, in good agreement with \citet{desert08}.  Because of the
range of temperatures found in a small spatial area, presumably all at
the same distance only 1.3\,kpc from the Sun, RCW~120 is an ideal
location to test for the $\beta-T$ relationship.  With the wavelength
coverage of {\it Herschel}, we are able to simultaneously constrain
$\beta$ and $T$ for regions of cold dust.

The temperature map of the RCW~120 field reveals numerous locations
in the PDR where radiation appears to be leaking into the
surrounding interstellar medium.  This implies that RCW~120 is in the
process of fragmenting its nearly complete PDR layer.  As the
radiation leaves the confined environment, the radiation
pressure shapes local cold filaments and may aid in collapsing local
condensations outside the ionization front.
 
\begin{acknowledgements}
Part of this work was supported by the ANR ({\it Agence Nationale pour
  la Recherche}) project ``PROBeS'', number ANR-08-BLAN-0241.  PACS
has been developed by a consortium of institutes led by MPE (Germany)
and including UVIE (Austria); KUL, CSL, IMEC (Belgium); CEA, LAM
(France); MPIA (Germany); IFSI, OAP/AOT, OAA/CAISMI, LENS, SISSA
(Italy); IAC (Spain). This development has been supported by the
funding agencies BMVIT (Austria), ESA-PRODEX (Belgium), CEA/CNES
(France), DLR (Germany), ASI (Italy), and CICT/MCT (Spain).  SPIRE has
been developed by a consortium of institutes led by Cardiff Univ. (UK)
and including Univ. Lethbridge (Canada); NAOC (China); CEA, LAM
(France); IFSI, Univ. Padua (Italy); IAC (Spain); Stockholm
Observatory (Sweden); Imperial College London, RAL, UCL-MSSL, UKATC,
Univ. Sussex (UK); Caltech, JPL, NHSC, Univ. Colorado (USA). This
development has been supported by national funding agencies: CSA
(Canada); NAOC (China); CEA, CNES, CNRS (France); ASI (Italy); MCINN
(Spain); SNSB (Sweden); STFC (UK); and NASA (USA).

\end{acknowledgements}

\bibliographystyle{14657} 
\bibliography{14657.bib}

\end{document}